\documentclass[a4paper]{jpconf}
\usepackage{graphicx}
\def \beq{\begin{equation}}
\def \eeq{\end{equation}}
\def \beqa{\begin{eqnarray}}
\def \eeqa{\end{eqnarray}}
\def \ie{{\sl i.e.\/}}
\def \rhov{\rho_{\scriptscriptstyle EM}}
\def \prop{G_{\scriptscriptstyle EM}}
\def \vecp{G_{\scriptscriptstyle V}}
\def \vertex{C_{\scriptscriptstyle EM}}
\begin{document}
\title{A transport coefficient: the electrical conductivity}
\author{Sourendu Gupta}
\address{Department of Theoretical Physics, Tata Institute of Fundamental Research,
   Homi Bhabha Road, Mumbai 400005, India.}

\ead{sgupta@tifr.res.in}

\begin{abstract}
I describe the lattice determination of the electrical conductivity
of the quark gluon plasma \cite{papers}. Since this is the first
extraction of a transport coefficient with a degree of control
over errors, I next use this to make estimates of other transport related
quantities using simple kinetic theory formul{\ae}. The resulting
estimates are applied to fluctuations, ultra-soft photon spectra
and the viscosity. Dimming of ultra-soft photons is exponential in
the mean free path, and hence is a very sensitive probe of transport.
\end{abstract}.

\section{Introduction}
Transport coefficients have recently been computed to leading
logarithmic order in weak coupling perturbation theory \cite{amy,amy2}.
However, at temeperatures of interest to heavy-ion collisions
experiments, the gauge coupling $g={\cal O}(1)$ and this computation
is ineffective. At the other extreme of $g\to\infty$, the AdS/CFT
correspondence has been exploited to bound the shear viscosity
$\eta/S\ge1/4\pi$ \cite{son}, where $S$ is the extropy density.
This result is for a highly supersymmetric version of QCD and needs
to be corrected in order to connect with experiments.

There is clearly a case for a lattice computation, if one can have
control over the process of analytic continuation from Euclidean
to Minkowski space-time. Direct computations of the shear viscosity
were tried \cite{karsch,nakamura}, but were inconclusive because
statistical problems beset even the Euclidean computations of the
correlator of the energy-momentum tensor, thus making it hard to
check the analytic continuation.

In this work \cite{papers} I shift the focus slightly and compute
the electrical conductivity, $\sigma$, of the quark-gluon plasma.
This involves a lattice computation of the correlator of the
electromagnetic current.  Since this is no harder than the extraction
of the $\rho$-meson mass, statistical errors in the Euclidean
computation are under control, and effort can be concentrated on
the harder job of analytical continuation. We are able to demonstrate
control over this process, and thereby extract $\sigma$.

This is interesting in itself for heavy-ion collisions because
several properties of soft photons in media can be inferred once
$\sigma$ is known. Since this is the first statistically reliable
extraction of a transport coefficient from lattice QCD, one can use
this to estimate many different quantities which depend on transport,
and are important for heavy-ion physics. With $\sigma$ in hand, and
using two well-specified approximations, we are able to predict the
value of $\eta$, and compare the lattice results with others.

The lattice computations described in Section 2 are from \cite{papers}.
The material in Section 3 is partly new.

\section{The lattice computation}

All lattice computations of transport coefficients are performed in
the context of linear response theory. The response, $A$, of a system
to a time-dependent force $F$ is given, in the frequency domain by
$A(\omega)=\chi(\omega)F(\omega)$. The response function $\chi(\omega)$
can then be written in terms of the retarded correlation function of $A$, and
the transport coefficient can be extracted using Kubo formul\ae.
For $\sigma$, the Kubo formula is
\beq
   \sigma(T) = \frac16\left.\frac{\partial}{\partial\omega}
      {\rhov}_i^i(\omega,{\mathbf 0},T)\right|_{\omega=0},
\label{cond}\eeq
where the sum is over spatial polarisations of the imaginary
part of the retarded photon propagator, {\sl i.e.\/}, the spectral
density, $\rhov^{\mu\nu}$, of the electromagnetic current correlator.
The usual QED Ward identity gives
$\rho^{00}(\omega,{\mathbf 0},T)=2\pi\chi_Q\omega\delta(\omega)=0$,
where $\chi_Q$ is the usual charge susceptibility.

A lattice computation proceeds from the spectral representation for
Euclidean current correlators---
\beq
   \prop(t,T) = \int_0^\infty \frac{d\omega}{2\pi}
       K(\omega,t,T) \rhov(\omega,T),
\label{spec}\eeq
where $K$ is the free propagator, $\prop$ is the product of the
zero (spatial) momentum vector correlator summed over all spatial
polarisations, $\vecp$, and the EM vertex factor $\vertex =
4\pi\alpha\sum_fe_f^2$, where $e_f$ is the charge of a quark of
flavour $f$. For later use we note that $\vertex\approx1/20$ for
two flavours. 

The main problem in this determination is that $\rhov$ is needed
at an infinite number of points and $\prop$ is known only at $N_t$
different values.  The solution is to constrain the function $\rhov$
through an informed guess, and use a Bayesian method to extract it.

We note a second complication. A free field theory computation shows
that $\rhov(\omega,T)$ increases as $\omega^3$ at large $\omega$.
This means that a naive application of the Kubo formul{\ae} would
violate causality \cite{hilge}. We try to solve this problem by
using $\Delta\prop(t,T)$ which is the difference of the measured
and free-field theory propagators to extract $\Delta\rhov$.

\begin{figure}[h]
\begin{minipage}{18pc}
\includegraphics[width=18pc]{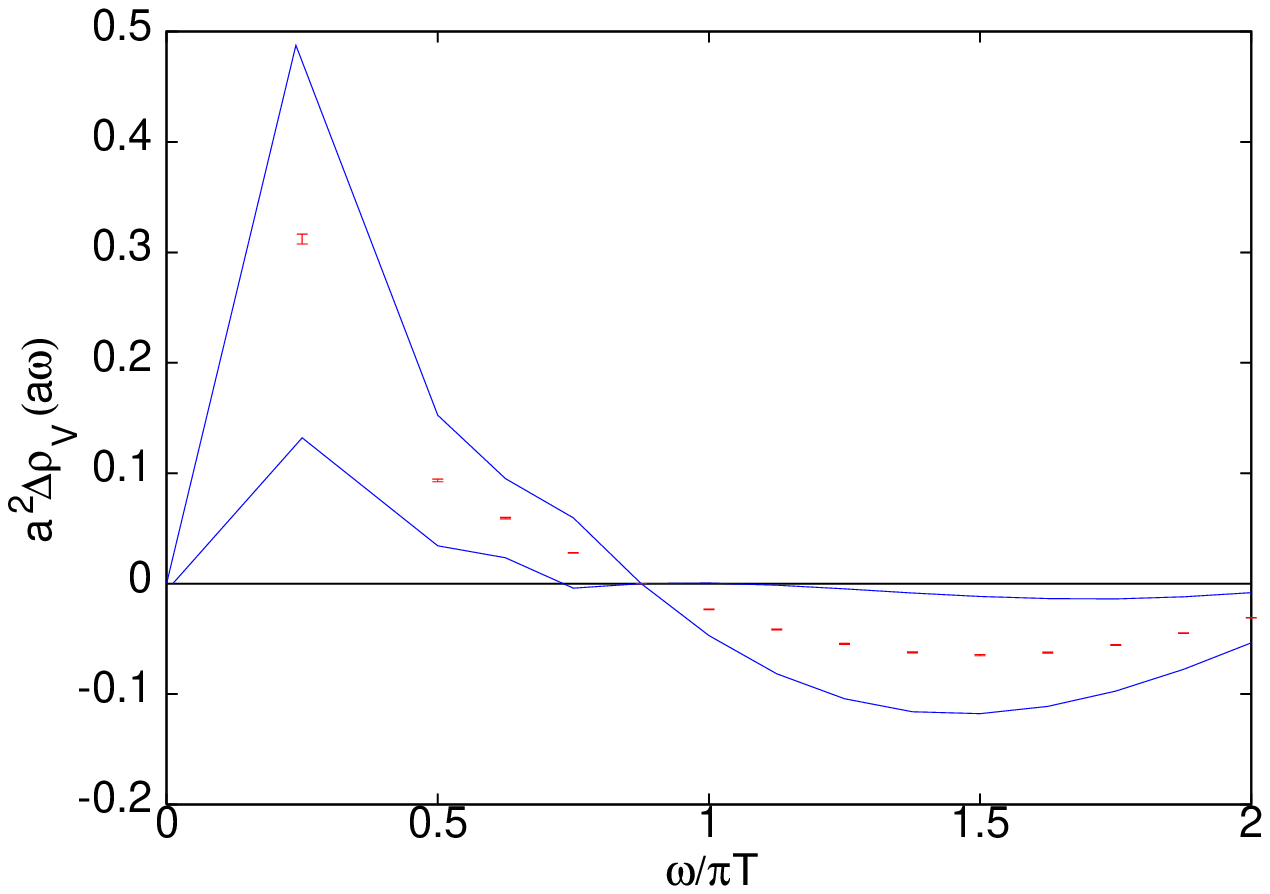}
\caption{\label{fg.rho}The spectral density $\Delta\rhov$ for $T=2T_c$ on
  a $12\times26^2\times48$ lattice with quark mass $0.03T_c$. See the text
  for a discussion of statistical and systematic errors.}
\end{minipage}\hspace{2pc}%
\begin{minipage}{18pc}
\includegraphics[width=18pc]{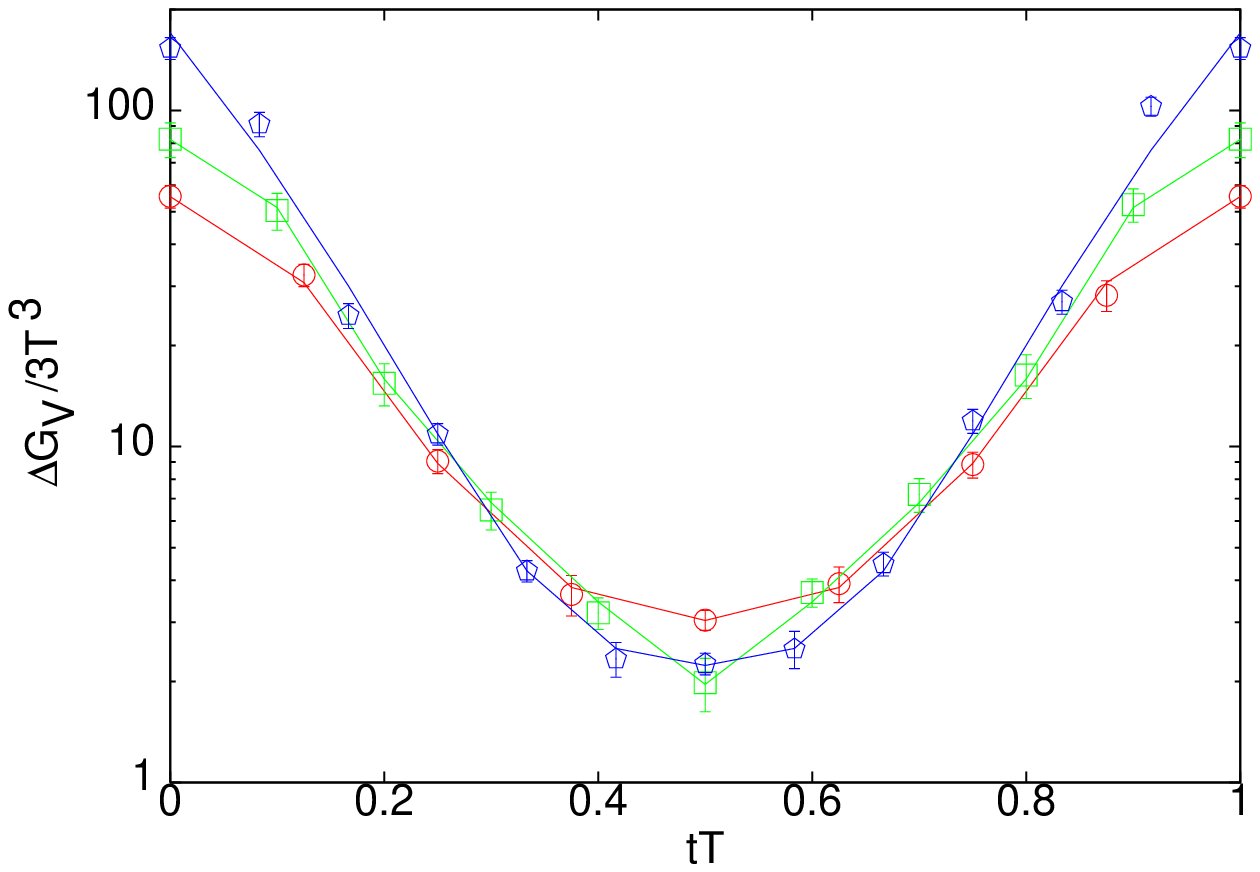}
\caption{\label{fg.prop}The propagator $\Delta\prop$ obtained through the
  fit of $\Delta\rhov$ compared to the data for $T=2T_c$ and $N_t=8$
  (circles), 10 (squares) and 12 (pentagons) with quark mass $0.03T_c$.}
\end{minipage} 
\end{figure}

Using a variety of Bayesian methods \cite{papers} to extract
$\Delta\prop(t,T)$ we found a peak at $\omega/(2\pi T)\approx0.15$
\cite{papers} and a smooth drop to zero at $\omega=0$ (see Figure
\ref{fg.rho}). The checks which have been performed on this extraction
are---
\begin{enumerate}
\item The statistical errors are negligible, as seen in Figures \ref{fg.rho}
    and \ref{fg.prop}.
\item The position of the peak seen in Figure \ref{fg.rho} is not strongly
    dependent on the lattice spacing, as checked by using $N_t=8$, 10 and 12.
\item Variations in the results are investigated due to algorithmic changes
    in the Bayesian prior and the truncation and discreitzation of
    eq.\ (\ref{spec}). The width of the band of systematic errors shown in
    Figure \ref{fg.rho} is thrice the width due to such variation.
\end{enumerate}
It would be useful to have similiar tests of robustness on all
uses of Bayesian methods including the maximum entropy method (MEM). 

The goodness of fit to the Euclidean data can be judged from Figure
\ref{fg.prop}. We draw attention to the small error bars on this
data when compared to the errors on typical measurements of Euclidean
correlators of energy momentum tensor.

Having established the existence of the peak, further information
is extracted using a parametrization of the peak in $\Delta\rhov$
as the ratio of two polynomials, with that in the numerator being
of lower order than the denonimator, so that the integrals used in
the Kramers-Kr\"onig relation converge. The denominator has to be
at least a quartic polynomial in order to smoothly connect to the
weak-coupling picture of a pinch singularity giving rise
to the peak along real $\omega$ \cite{aarts}. (We note in this
connection that earlier attempts to extract transport quantities
have worked with the full $\rhov$ rather than with $\Delta\rhov$.
This error can strongly suppress estimates of the transport
coefficients. Since $\rhov$ increases with $\omega$, this biases
the fitted peak towards larger $\omega$, thus decreasing the slope
at the origin for any fit form with one peak.) Using these
fits our estimate of the electrical conductivity is--- \beq
   \frac{\sigma(T)}T = \vertex\times
      \cases{7.5\pm0.8 & ($T=1.5T_c$)\cr
             7.7\pm0.6 & ($T=2T_c$)\cr
             7.0\pm0.4 & ($T=3T_c$)}
\label{result}\eeq

\section{Appplications}

\subsection{Viscosity}
In kinetic theory one can write the Drude formula for the electrical
conductivity---
\beq
   \sigma = \frac{\vertex S_q\tau_q}m,
\label{transport}\eeq
where $S_q$ is the entropy density in quarks, $m$ their screening
mass and $\tau_q$ their mean free time. Using the lattice result
for $\sigma$ one gets $\tau_q=0.30\pm0.03$ fm.  While charge is
transported through quarks, momentum is transported mainly through
gluons.  Due to colour factors the mean free time for gluons is
about half the mean free time for quarks. Then using a kinetic
theory formula for the viscosity along with such an estimate of
$\tau_g$ one gets
\beq
   \frac\eta S\approx0.21\pm0.02 > \frac1{4\pi}\,.
\label{visco}\eeq
Only lattice errors are shown in this estimate. Errors on the kinetic
theory are harder to extract. The lattice + kinetic theory result
is 2.5 times the AdS/CFT limit.

\subsection{Photon dimming}
Non-zero electrical conductivity leads to screening of photons with
energy below the peak of the spectral function found earlier, \ie,
for $\omega<300$ MeV. This is a direct consequence of Maxwell's
equations. This range of screened energies lies far below the excess
which may have been seen by the CERN NA49 experiment.
It seems possible that direct photons at such low-energy could be
distinguished from photons coming from decays of pions at RHIC-II.
One method for using an observation of photon dimming to extract
the conductivity has been suggested \cite{photon}.

The mean free path, $\ell$, of these ultra-soft photons is given by
the formula
\beq
   \ell=\frac{\tau_q}{\vertex}\approx6{\rm\ fm}.
\label{skin}\eeq
Since the fireball at RHIC is about 7 fm in size, one could expect
dimming by a factor of $\exp(-7/6)$, \ie, by a factor of 3.2.  There
is a self-consistency argument here involving only the data. If the
viscosity is small, then mean free times of quarks and gluons are
small, and the mean free path of photons can only be about 20
($=1/\vertex$) times larger.  However, dimming is exponential in
the mean free path. For example, the mean free path as predicted
in AdS/CFT would lead to dimming by a factor of 18.8.  Thus, the
use of photon dimming to extract $\sigma$ is an important cross
check of values of $\eta$ extracted by other means.

\subsection{Fluctuations}
The study of event-to-event fluctuations in limited rapidity bins is
expected to give information on the phase diagram of QCD as well as
make contact between experiments and basic lattice QCD computations.
However, fluctuations of conserved quantities in a limited region of
space are expected to even out by transport processes such as diffusion.
The computation of $\sigma$ is relevant not only to the diffusion of
charge, but also of baryon number and isospin.

The diffusion constant can be obtained using the kinetic
theory formula \cite{gavin}
\beq
   D=\tau_qc_s^2\approx0.1{\rm\ fm},
\label{einstein}\eeq
where we have taken the square of the speed of sound $c_s^2=0.30\pm0.01$
at $T=2T_c$, as shown by recent lattice computations \cite{swagato}.
If the plasma is highly dissipative, then $D$ must be small.
When $D$ decreases, the
fluctuation are more localized at the same time.
Small viscosity implies smaller mean free path,
and hence longer persistence of the fluctuation signal. In the
approach of \cite{gavin} this would mean that the rapidity width
of a fluctuation is $\Delta Y\le2D/\tau_0\approx0.4$ for fluctuations
created at time $\tau_0=0.5$ fm.

\subsection{A brave new world}
When weak coupling was the only tool to investigate transport
properties, it was felt that the QGP would be weakly dissipative.
Now with other tools at our disposal, the question seems more open,
and preliminary evidence is that dissipation is strong. The first
lattice results on transport \cite{papers} pin down many aspects
of transport, as outlined in this section, and lead to the hope
that multiple experimental constraints on transport are feasible.

\end{document}